\documentclass[
    ,final            % use final for the camera ready runs
  ]
  {aipproc}

\layoutstyle{8x11single}

\usepackage{bm}        % for math
\usepackage{amssymb}   % for math
\usepackage{amsmath}   % for math
\usepackage{tabularx}   % for math
\begin{document}
\title{Lattice NRQCD study on in-medium bottomonium
   spectra using a novel Bayesian reconstruction approach}


\author{Seyong Kim}{
  address={Department of Physics, Sejong University, Seoul 143-747, South Korea}
}

\author{Peter Petreczky}{
  address={Physics Department, Brookhaven National Laboratory, Upton, NY 11973, USA}
}

\author{Alexander Rothkopf}{
  address={Institute for Theoretical Physics,  Heidelberg University, Philosophenweg 16, 69120 Heidelberg, Germany}
}

\begin{abstract}
We present recent results on the in-medium modification of S- and P-wave bottomonium states around the deconfinement transition. Our study uses lattice QCD with $N_f=2+1$ light quark flavors to describe the non-perturbative thermal QCD medium between $140$MeV$<T<249$MeV and deploys lattice regularized non-relativistic QCD (NRQCD) effective field theory to capture the physics of heavy quark bound states immersed therein. The spectral functions of the $^3S_1$ $(\Upsilon)$ and $^3P_1$ $(\chi_{b1})$ bottomonium states are extracted from Euclidean time Monte Carlo simulations using a novel Bayesian prescription, which provides higher accuracy than the Maximum Entropy Method. Based on a systematic comparison of interacting and free spectral functions we conclude that the ground states of both the S-wave $(\Upsilon)$ and P-wave $(\chi_{b1})$ channel survive up to $T=249$MeV. Stringent upper limits on the size of the in-medium modification of bottomonium masses and widths are provided.
\end{abstract}

\maketitle

%%%%%%%%%%%%%%%%%%%%%%%%%%%%%%%%%%%%%%%%%%%%
%% MAINMATTER
%%%%%%%%%%%%%%%%%%%%%%%%%%%%%%%%%%%%%%%%%%%%

\section{Introduction}
 
The fact that the strong interactions are asymptotically free tells us that the partons confined in hadrons at everyday energy scales will eventually be liberated at high energies. Nuclear matter under extreme conditions provides a fascinating laboratory to elucidate the transition between these regimes of confinement and deconfinement. The strong thermal fluctuations present shortly after the big bang or the high densities in the interior of neutron stars indeed lead to the emergence of partons as the relevant degrees of freedom of a strongly correlated quark-gluon plasma (QGP). Modern collider facilities, such as RHIC and LHC and future machines, such as FAIR, provide a direct window to these scenarios in the laboratory. At the same time lattice QCD simulations at finite temperature \cite{Borsanyi:2013bia, Bazavov:2014pvz} (and more recently at finite density \cite{Sexty:2013ica}) give theory non-perturbative access to the properties of a realistic QCD medium. The strength of this approach lies in being valid both in the confined and the deconfined phase, as well as close to the transition, which is out of reach of perturbation theory.

To extract insight from experimental measurements and theory simulations, we need to identify appropriate physics probes. One system particularly suited to investigate aspects of deconfinement are the bound states of a heavy quark and anti-quark, heavy quarkonium \cite{Brambilla:2010cs}. In vacuum, the lifetime of the ground states of these two-body systems is of the order of inverse keV \cite{Agashe:2014kda}, rendering them essentially stable on nuclear time scales. Surrounded by a thermal medium these states are susceptible to the effects of thermal fluctuations but remain well defined particles that can be cleanly identified in experiment \cite{Mocsy:2013syh}. As temperature increases, the binding forces weaken and eventually the heavy mesons will also melt into their constituents. The question at which temperature this melting occurs is an active area of research in the lattice QCD community (see e.g. \cite{Umeda:2002vr,Aarts:2010ek}).

In the context of relativistic heavy-ion collisions, the melting of heavy quarkonium plays a central role. The related suppression of its yields was initially proposed as the prime signal \cite{Matsui:1986dk} for the creation of a quark-gluon plasma in the collision center. Recent precision measurements of bottomonium spectra (the $b\bar{b}$ vector mesons) at RHIC and LHC \cite{Chatrchyan:2012lxa} have indeed revealed a pattern of sequential suppression \cite{Karsch:1987pv}. A first principles understanding of the survival of individual bound states from lattice QCD hence promises to establish heavy quarkonium as a QCD thermometer of the QGP.

While standard lattice QCD is well suited to describe the light medium degrees of freedom at finite temperature, it is challenging to directly include heavy quarks. In order to resolve a quark of mass $M_Q$, the lattice spacing must be chosen fine enough $aM_Q\ll1$ to suppress discretization artifacts. For bottomonium $M_b=4.66$GeV it is feasible to generate such lattices in the absence of light medium quarks (quenched approximation) \cite{Datta:2006ua}, but it is prohibitively costly in full QCD. The separation of scales between the constituent mass and those of the surrounding medium, which underlies this issue can however be turned into an advantage. It allows us to go over to an effective field theory (EFT) description called non-relativistic QCD (NRQCD) \cite{Brambilla:2004jw}. In it the heavy quarks are not part of the thermal bath but propagate as probes in the background of the medium gauge fields, which are simulated by standard lattice QCD methods.

Lattice regularized NRQCD \cite{Thacker:1990bm} is a well established approach for the study of the vacuum properties of bottomonium (see e.g. \cite{Gray:2005ur}). Its application at finite temperature began with \cite{Fingberg:1997qd} and has recently been invigorated through a series of studies of bottomonium in a thermal medium with $2$ and $2+1$ light quark flavors \cite{Aarts:2010ek,Aarts:2014cda}. Unfortunately the pion mass on the deployed lattices remained rather large at around $400$MeV. Our study \cite{Kim:2014iga} relies on current generation lattices with $N_f=2+1$ quark flavors and Highly Improved Staggered Quark (HISQ) action generated by the HotQCD collaboration \cite{Bazavov:2011nk}. The pion mass in the continuum limit $M_\pi\simeq 161$MeV on these isotropic $48\times12$ configurations lies close to the physical value and the chiral transition is located at $T_c=159$MeV.  

The determination of spectral information from lattice QCD data represents a significant challenge. Monte Carlo simulations are performed in Euclidean time, i.e. dynamical quantities, such as spectral functions, are not directly accessible. In practice we face the task to extract from a finite and noisy set of numerical correlator data a continuous function encoding the intricate spectrum of in-medium bottomonium. This constitutes an inherently ill-defined problem, as a simple $\chi^2$ fit to the data would lead to an infinite number of degenerate spectra. It can however been given meaning by the use of Bayesian inference \cite{Jarrell:1996}. This well established statistical approach provides the mathematical framework to use prior knowledge, such as the positivity or smoothness of the spectrum, in order to regularize the otherwise under-determined $\chi^2$ fit. The Maximum Entropy Method (MEM) \cite{Asakawa:2000tr} used in the previous studies on in-medium bottomonium represents one possible choice of regularization for this problem. 

In the following we will we deploy a recently developed Bayesian approach \cite{Burnier:2013nla} that differs significantly from the MEM. While the latter uses arguments from image reconstruction \cite{Jarrell:1996}, the novel method is based on three conditions directly related to the problem at hand: (1) It enforces the positivity of the spectrum, (2) it guarantees that the end result is independent from the choice of units for $\rho(\omega)$ , and (3) that the reconstructed spectrum is smooth except in the frequency region where data encodes peaked structures. From these conditions we obtain a regularization functional different from the Shannon-Jaynes Entropy used in the MEM. In particular it is devoid of the flat directions that hamper the convergence of the MEM. (See \cite{Burnier:2013nla} for technical details and \cite{Burnier:2014ssa} for another application of this new method). 

\section{Results at Zero Temperature}

Bottomonium correlation functions in Euclidean time are the starting point of our study. They are constructed from an appropriate combination of bottom quark propagators and vertex operators that encode the quantum numbers of the bottomonium channel under investigation. The heavy quark evolution in the background of a thermal medium follows from the lattice discretized ${\cal O}(v^4)$ NRQCD Lagrangian \cite{Thacker:1990bm} and represents an initial value problem in imaginary time. The influence of temperature only enters through the coupling to the medium gauge fields, since periodic boundary conditions are not imposed on the quarkonium correlator. As an effective field theory, lattice NRQCD does not posses a naive continuum limit. This requires, in contrast to relativistic formulations, that the lattice spacing must be coarse enough to fulfill $aM_b>1.5$ \cite{Lepage:1992xa}. All HotQCD configurations deployed in the following obey this requirement. (See appendix B in \cite{Kim:2014iga}  and \cite{Kim:2014nda} for a discussion and numerical tests of the applicability of the NRQCD approximation.)

The process of setting up NRQCD involves integrating out the scale of the heavy quark rest mass \cite{Brambilla:2004jw}. In practice this amounts to the presence of a lattice spacing dependent shift $C(\beta)$ in the heavy quarkonium mass, measured in the EFT. In order to recover the absolute mass scale, one needs to perform a calibration at zero temperature in addition to the determination of the physical value of the lattice spacing. $C(\beta)$ does not depend on the bottomonium channel and we extract it from a comparison of the numerical $T=0$ mass $E^{\rm NRQCD}_{\Upsilon(1S)}$ of the S-wave ground state $\Upsilon(1S)$ with its experimentally determined value $M^{\rm exp}_{\Upsilon(1S)}$
\begin{align}
 M^{\rm exp}_{\Upsilon(1S)}=E^{\rm NRQCD}_{\Upsilon(1S)}+C(\beta).
\end{align}

\begin{figure}[t]
\centering
 \includegraphics[scale=0.3, angle=-90]{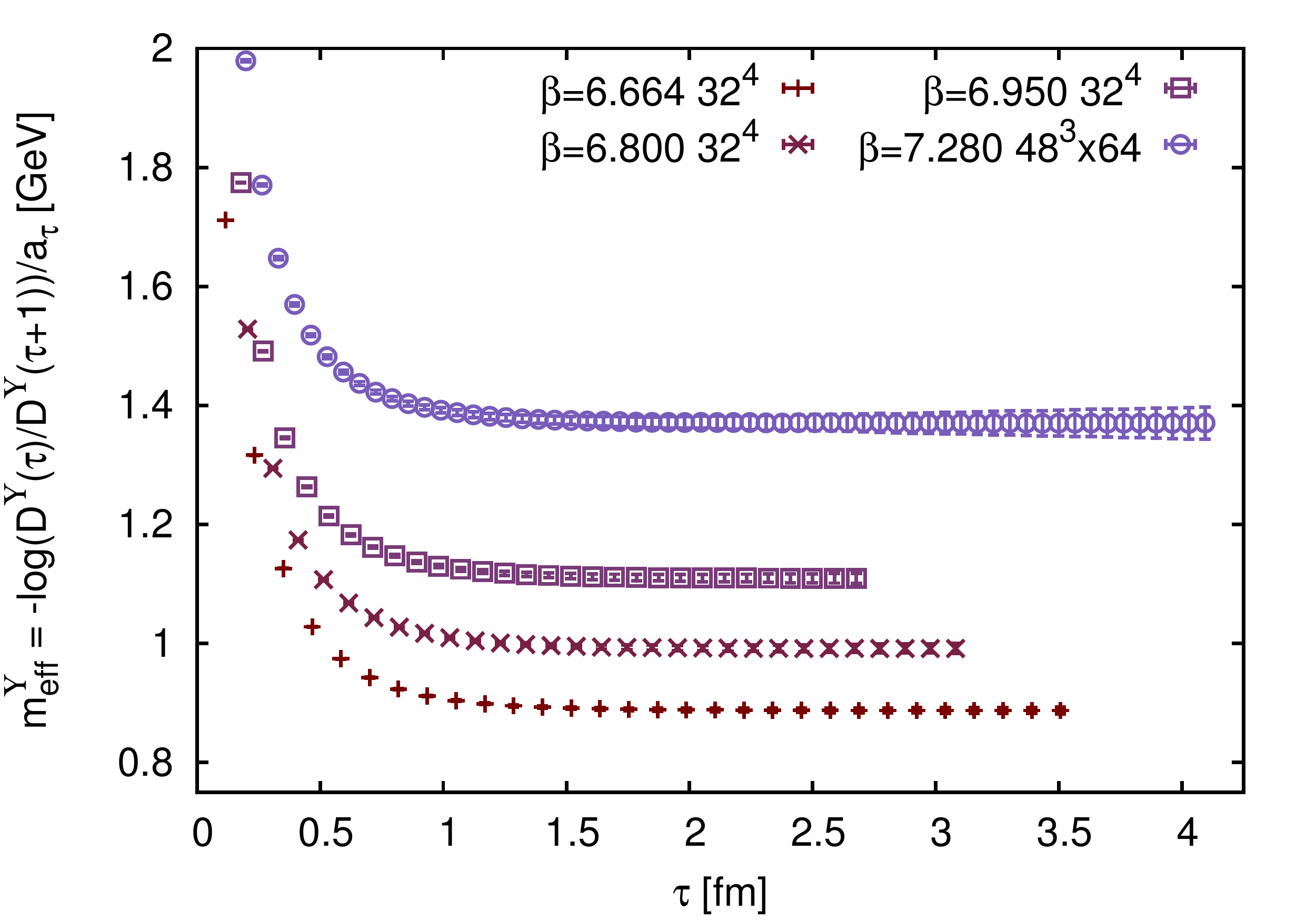} 
 \includegraphics[scale=0.3, angle=-90]{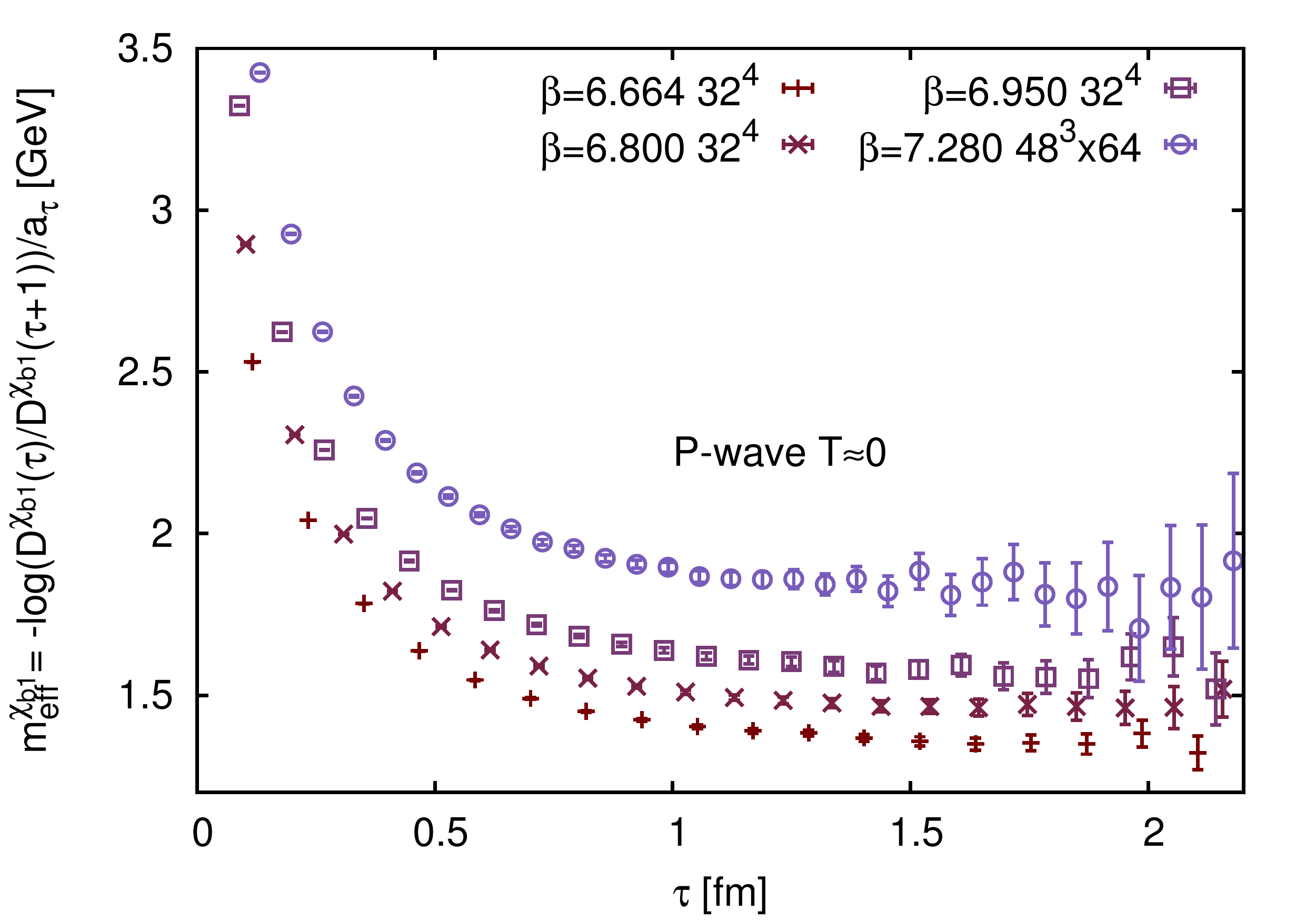} 
 \caption{Effective mass plot $m_{\rm eff}(\tau)$ for the S-wave (left) and P-wave (right) channel at $T\simeq0$. The four curves correspond to the lattice spacings $\beta=6.664,6.8,6.95$ and $7.28$. Due to the smaller signal-to noise ratio, we plot the P-wave only up to times of $\tau=2.2$fm. }\label{Fig1}
\end{figure}

Fig.\ref{Fig1} shows the effective mass $am_{\rm eff}(\tau)={\rm log}[D(\tau/a)/D(\tau/(a+1))]$ at zero temperature for the $\Upsilon$ (left) and $\chi_{b1}$ (right) channel. The four curves are obtained from lattices with gauge coupling $\beta=6.664, 6.8, 6.95$ and $7.28$. We see that indeed the value of the plateau at large $\tau$, which represents the same physical ground state, varies with $\beta$. Once $C(\beta)$ is fixed, we can use it to calibrate the mass of the P-wave ground state $\chi_{b1}$, which yields $M^{\rm NRQCD}_{\chi b1}=9.917(3)$GeV. This is slightly above the PDG value of $M_{\chi b1}(1P)=9.899278(26)(31)$GeV but compatible with $M^{\rm NRQCD}_{\chi b1}=9.921(15)$GeV obtained by another recent NRQCD study, also based on $N_f=2+1$ flavor lattices \cite{Aarts:2014cda}. Note that the correlator data for the P-wave channel is more noisy than the S-wave, which is due to the fact that the masses of the $\chi_{b1}$ ground state lies above those of Upsilon. With a stronger exponential suppression of the correlator but the same available statistics, the signal to noise ratio diminishes.

\begin{figure}[t]
\centering
 \includegraphics[scale=0.3, angle=-90]{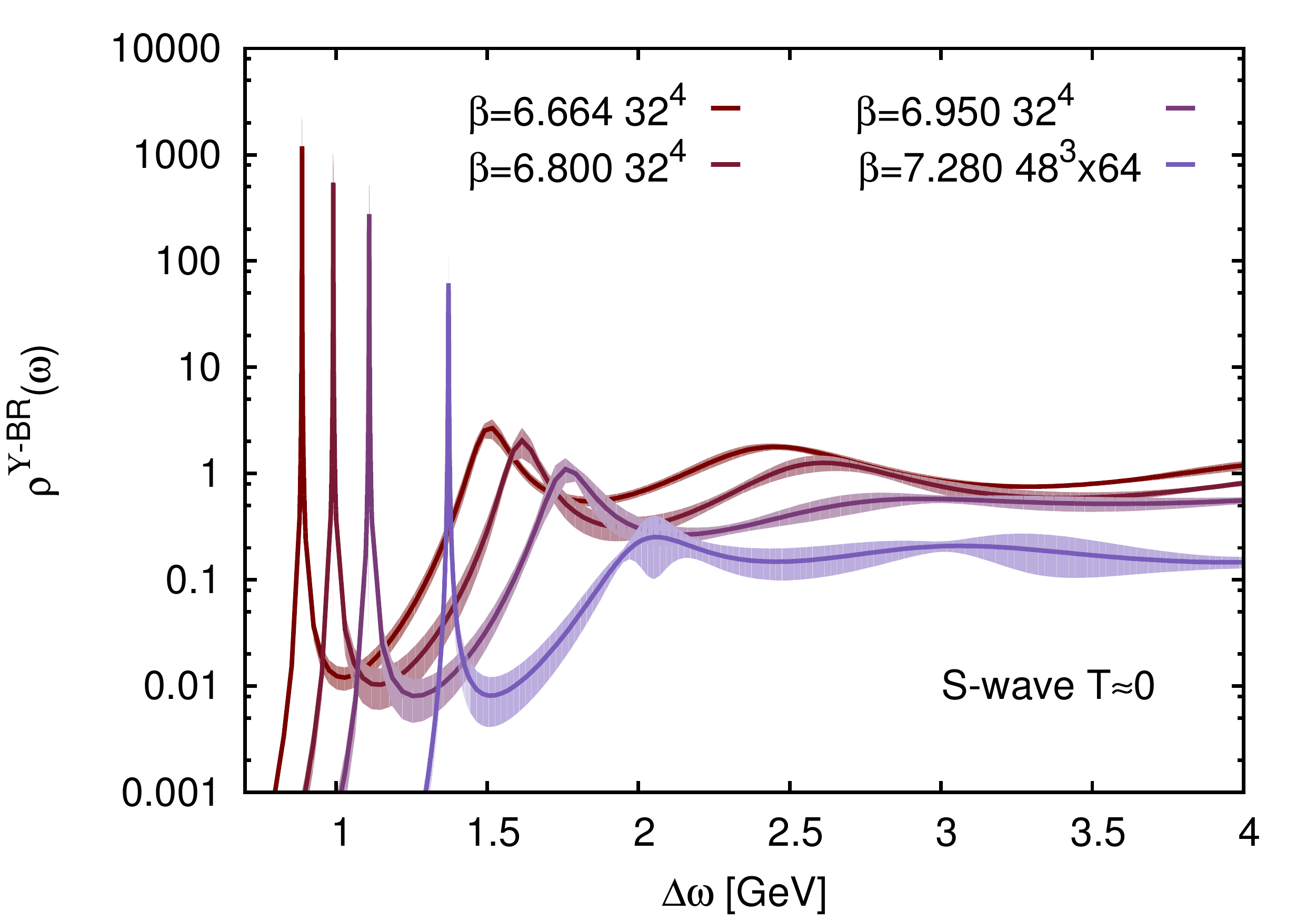} 
 \includegraphics[scale=0.3, angle=-90]{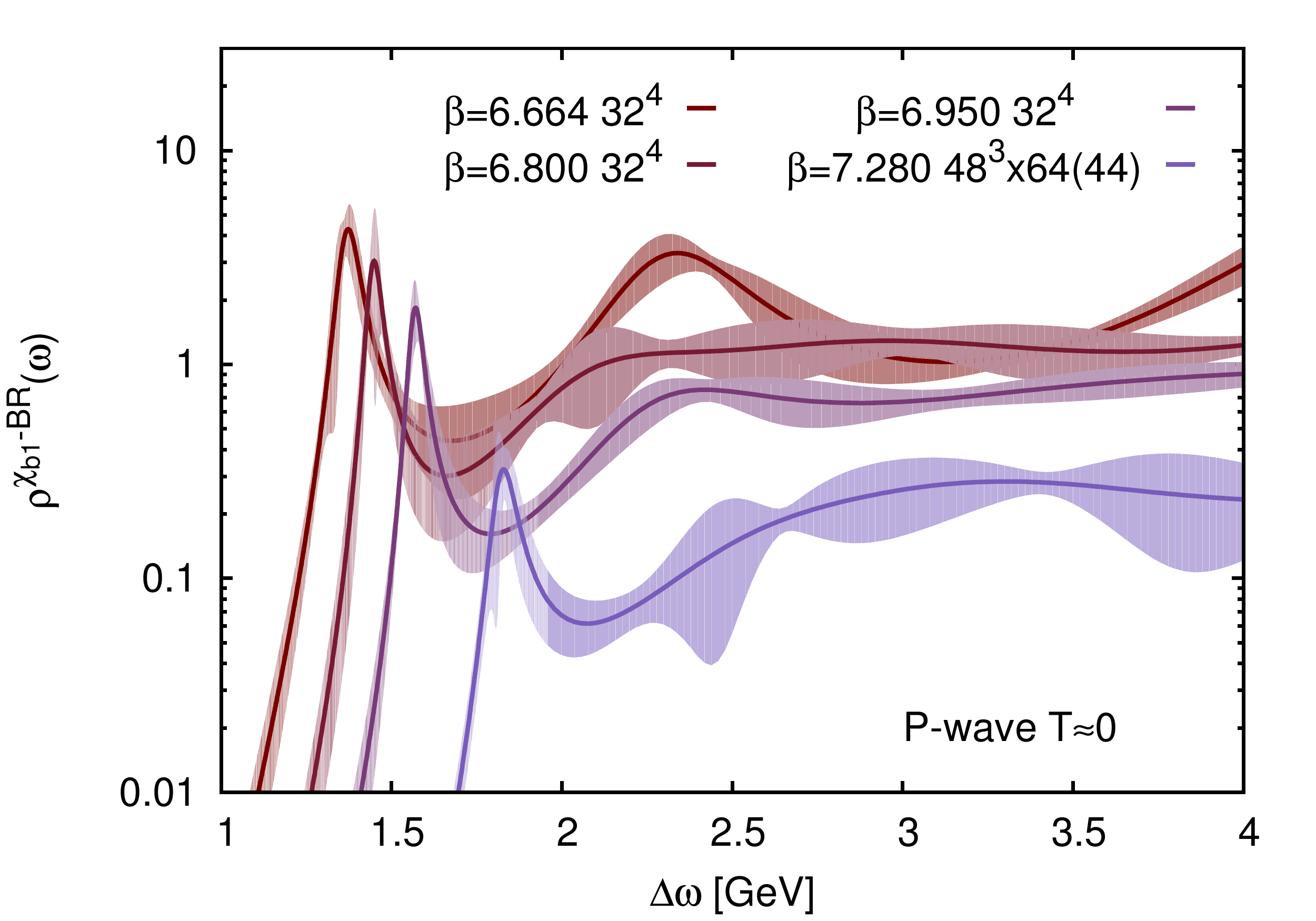} 
 \caption{Spectral functions in the S-wave and P-wave channel at $T\simeq0$ from a novel Bayesian reconstruction approach. The four curves correspond to the lattice spacings $\beta=6.664,6.8,6.95$ and $7.28$. The reconstruction of the P-wave spectrum at $\beta=7.280$ only takes into account data up to $\tau/a=44$, beyond which the signal in the correlator is lost. }\label{Fig2}
\end{figure}

We proceed by extracting the spectral functions from Euclidean correlator data based on a novel Bayesian prescription (for details and explicit parameters see \cite{Kim:2014iga}). One benefit of the NRQCD approach at finite temperature is that the relation between spectrum and correlator takes on the same form as at zero temperature in a relativistic formulation. The absence of periodicity in Euclidean time allows us to not only use all of the measured datapoints instead of just one-half but also eliminates the difficulties associated with the so called constant contribution 
problem \cite{Umeda:2007hy,Petreczky:2008px}.
Fig.\ref{Fig1} shows the results for the $\Upsilon$ (left) and $\chi_{b1}$ (right) channel. The error bands represent statistical errors. They are obtained from the variance between the outcome of ten reconstructions, in which consecutive blocks of $1/10$th of the available correlator measurements are disregarded. We find that the ground state of the Upsilon channel is very well resolved. Fits to its position with a Lorentzian give compatible values and comparable error bars, as found from the effective mass plot of Fig.\ref{Fig1}. The second peaked structure contains contributions from the $\Upsilon(2S)$ and higher excited states, the wiggles at higher frequencies are numerical artifacts. 

The P-wave channel also shows well defined ground state peaks, which however have a larger width and error bars than the S-wave. Besides the lower signal-to noise ratio in the correlator, it is the smaller physical peak size compared to the continuum contribution that makes the reconstruction more challenging here. 

\begin{figure}[t]
\centering
 \includegraphics[scale=0.3, angle=-90]{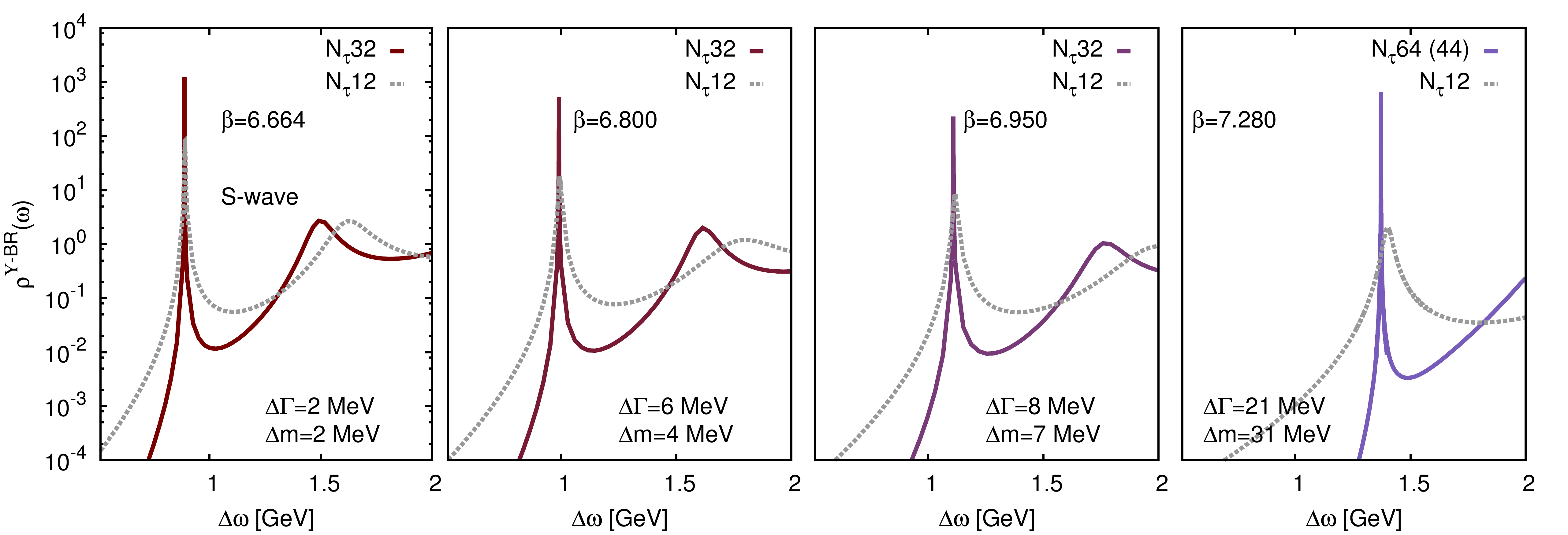}\newline
 \caption{Accuracy test for the S-wave channel: Comparison of the Bayesian spectral reconstructions of the full $T=0$ correlator dataset (solid line) with the result from truncating the same dataset at $\tau/a=12$ (gray dashed line). Note that that NRQCD is most reliable on the coarsest lattices $\beta=6.664$. The susceptibility of the reconstruction to the truncation grows as the lattice become more finely spaced.}\label{Fig3}
\end{figure}
\begin{figure}[t]
\centering
 \includegraphics[scale=0.3, angle=-90]{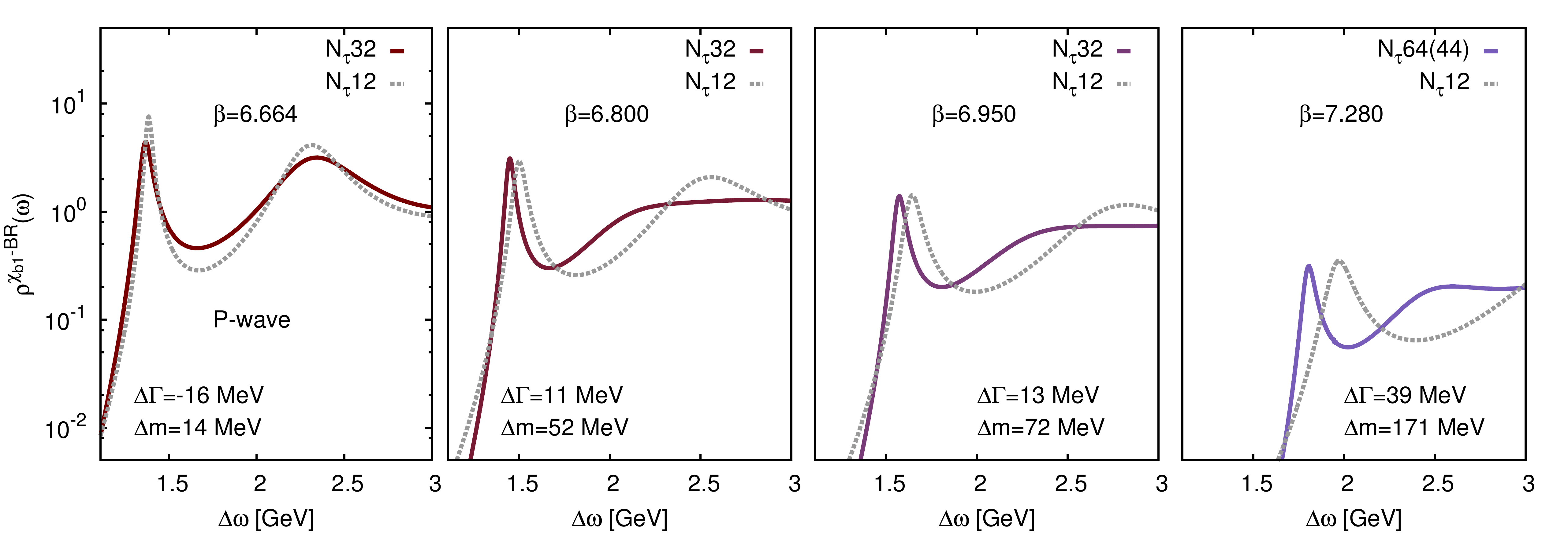} 
 \caption{Accuracy test for the P-wave channel: Comparison of the Bayesian spectral reconstructions of the full $T=0$ correlator dataset (solid line) with the result from truncating the same dataset at $\tau/a=12$ (gray dashed line). Due to the smaller signal-to-noise ratio and the smaller physical size of the peak compared to the continuum contribution the reconstructions suffer more strongly from the truncation.}\label{Fig4}
\end{figure}

If we wish to investigate the in-medium behavior of bottomonium in the following we have to make sure that we understand the limitations of the spectral reconstruction. Going from zero temperature to finite values of T is achieved by simulating on lattices with the same lattice spacing but a reduced number of points in temporal direction. This means that for the spectral reconstruction fewer datapoints will be available, which adversely affects the accuracy of the results. One possible test to quantify these effects is to repeat the spectral reconstruction based on the $T=0$ correlator dataset, truncated to the same number of datapoints $\tau_{\rm max}/a=12$ available at finite T. Fig.\ref{Fig3} and Fig\ref{Fig4} show the outcome for the S-wave and P-wave channel respectively. On the coarsest lattices, where NRQCD works best, the change in lowest peak position and width is rather small. The finer the lattice spacing the more susceptible the reconstruction becomes to a loss in $N_\tau$. Based on  these test (and additional checks laid out in \cite{Kim:2014iga}) we can determine stringent accuracy limits for mass shifts and width broadening of the S-wave 
\begin{align}
\nonumber \Delta m_{\Upsilon(1S)}(140{\rm MeV})&<2{\rm MeV}, \quad \Delta \Gamma_{\Upsilon(1S)}(140{\rm MeV})<5{\rm MeV}\\
 \Delta m_{\Upsilon(1S)}(249{\rm MeV})&<40{\rm MeV}, \quad \Delta \Gamma_{\Upsilon(1S)}(249{\rm MeV})<21{\rm MeV}.\label{Eq:SwaveBounds}
\end{align}
and P-wave spectra 
\begin{align} \nonumber |\Delta
  m_{\chi_{b1}(1P)}|(140{\rm MeV})&<60{\rm MeV}, \quad \Delta \Gamma_{\chi_{b1}(1P)}(140{\rm MeV})<20{\rm MeV}\\ \Delta m_{\chi_{b1}(1P)}(249{\rm
    MeV})&<200{\rm MeV}, \quad \Delta \Gamma_{\chi_{b1}(1P)}(249{\rm
    MeV})<40{\rm MeV}.\label{Eq:PwaveBounds}
\end{align} 
at the lowest $(140{\rm MeV})$ and highest $(249{\rm MeV})$ temperature. I.e. only changes in the finite temperature reconstruction that go beyond these values can be attributed to the effects of a thermal medium. 

\section{Results at Finite Temperature}

We proceed to present the main results of our study, the behavior of bottomonium at finite temperature. In addition to the four lattice spacing, for which also $T=0$ configurations were generated, we use a total of fourteen different values to achieve a fine temperature scan between $140$MeV$<T<249$MeV. The behavior of the in-medium correlators already tells us that while we can expect statistically significant in-medium effects at temperatures above the transition temperature, the absolute magnitude of the effects will be small. I.e. the ratio of the $T=0$ and the $T>0$ correlator differs at most by 1\% for the S-wave and at most by 5\% for the P-wave channel. This difference is related to the fact that the larger spatial extend of the $\chi_{b1}$ state makes it more susceptible to medium effects.

\begin{figure}[t!]
\centering
 \includegraphics[scale=0.3, angle=-90]{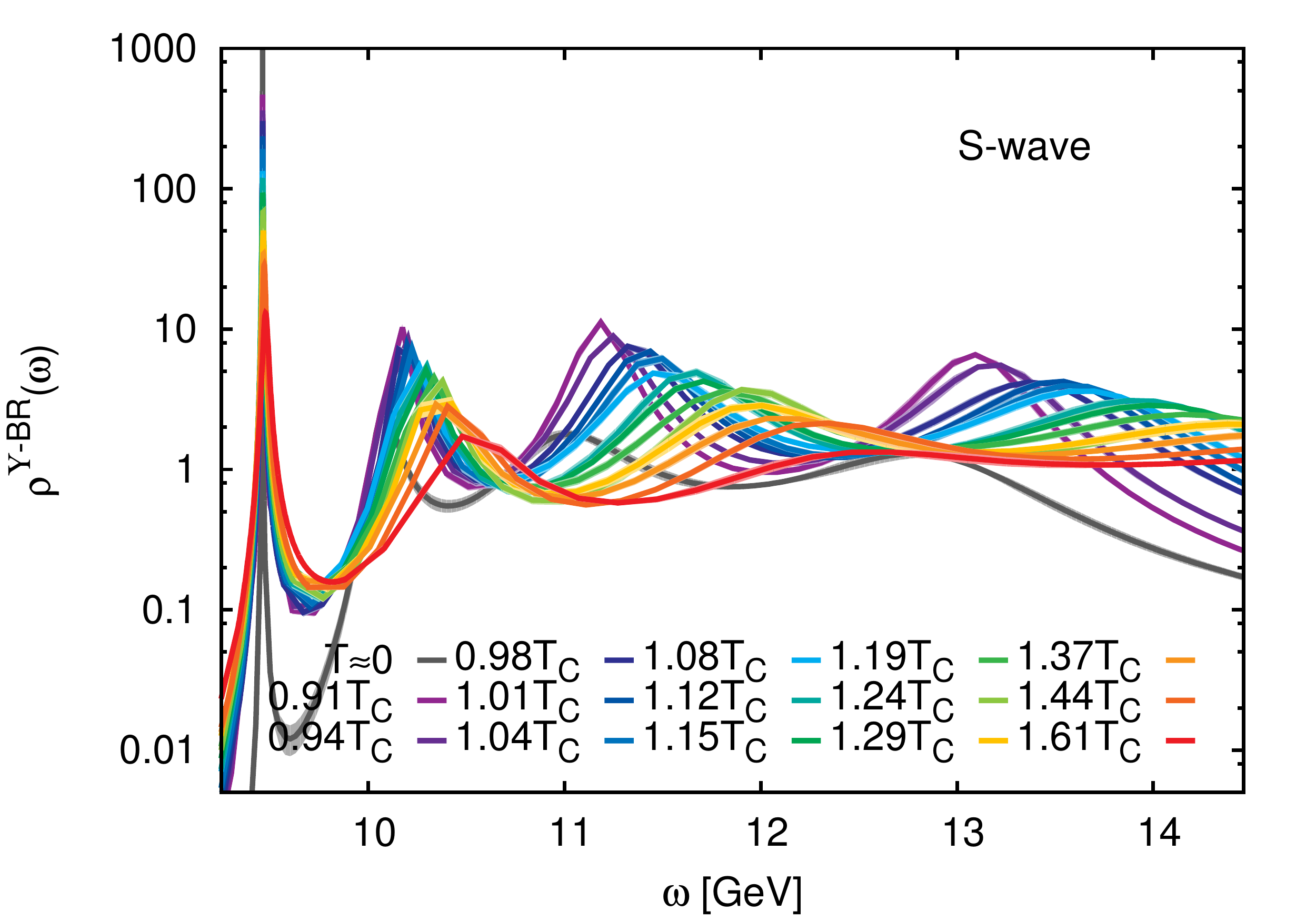}
 \includegraphics[scale=0.3, angle=-90]{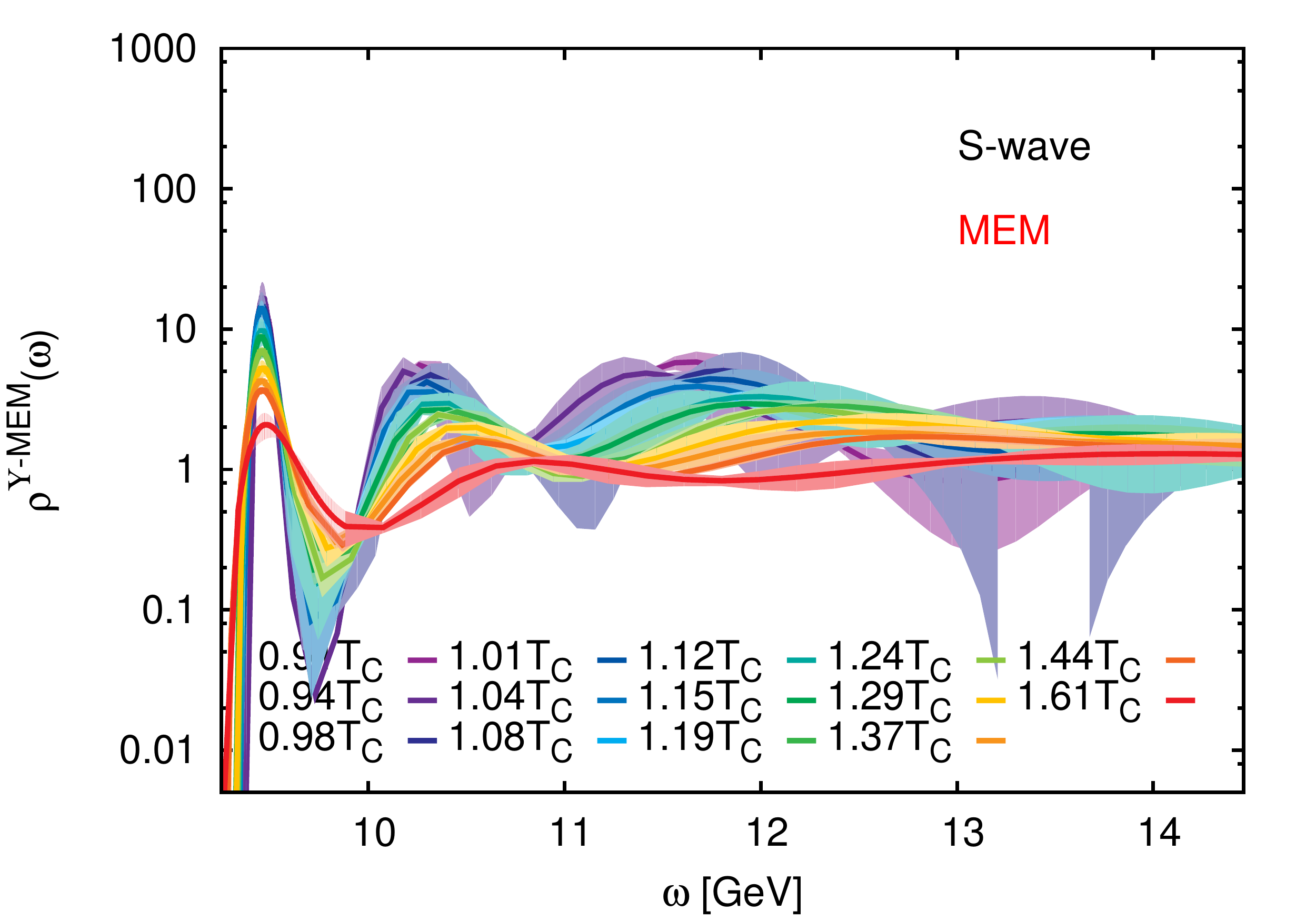}
 \caption{S-wave bottomonium spectral functions at finite temperature from the novel Bayesian approach (left) and the MEM (right). The fourteen $T>0$ curves span the range of $140$MeV$<T<249$MeV. The Bayesian reconstruction shows a well resolved and Lorentzian shaped ground state peak at all temperatures. The spectral features from the MEM, while qualitatively consistent are more shallow.}\label{Fig5}
\end{figure}

In Fig.\ref{Fig5} we show the results for the S-wave channel spectra, obtained from the novel Bayesian approach (left), which are contrasted to the MEM (right) results based on the same dataset. The values of $C(\beta)$, determined at $T=0$, are used to set the absolute frequency scale.

Both methods give qualitatively consistent results, i.e. a well defined ground state peak is visible up to the highest available temperature of $249$MeV. Note however that the reconstructed features of the MEM are much more shallow and that the ground state peak shape resembles a Gaussian. For a particle of finite lifetime we however expect to find a Lorentzian, which is what the novel Bayesian approach actually shows. A small mass shift and width broadening can be seen in the left panel of Fig.\ref{Fig5}, the strength of these effects however is smaller than the accuracy limits of the reconstruction, which thus remain as stringent upper limits to the in-medium modification of the S-wave channel. 

\begin{figure}[t!]
\centering
 \includegraphics[scale=0.3, angle=-90]{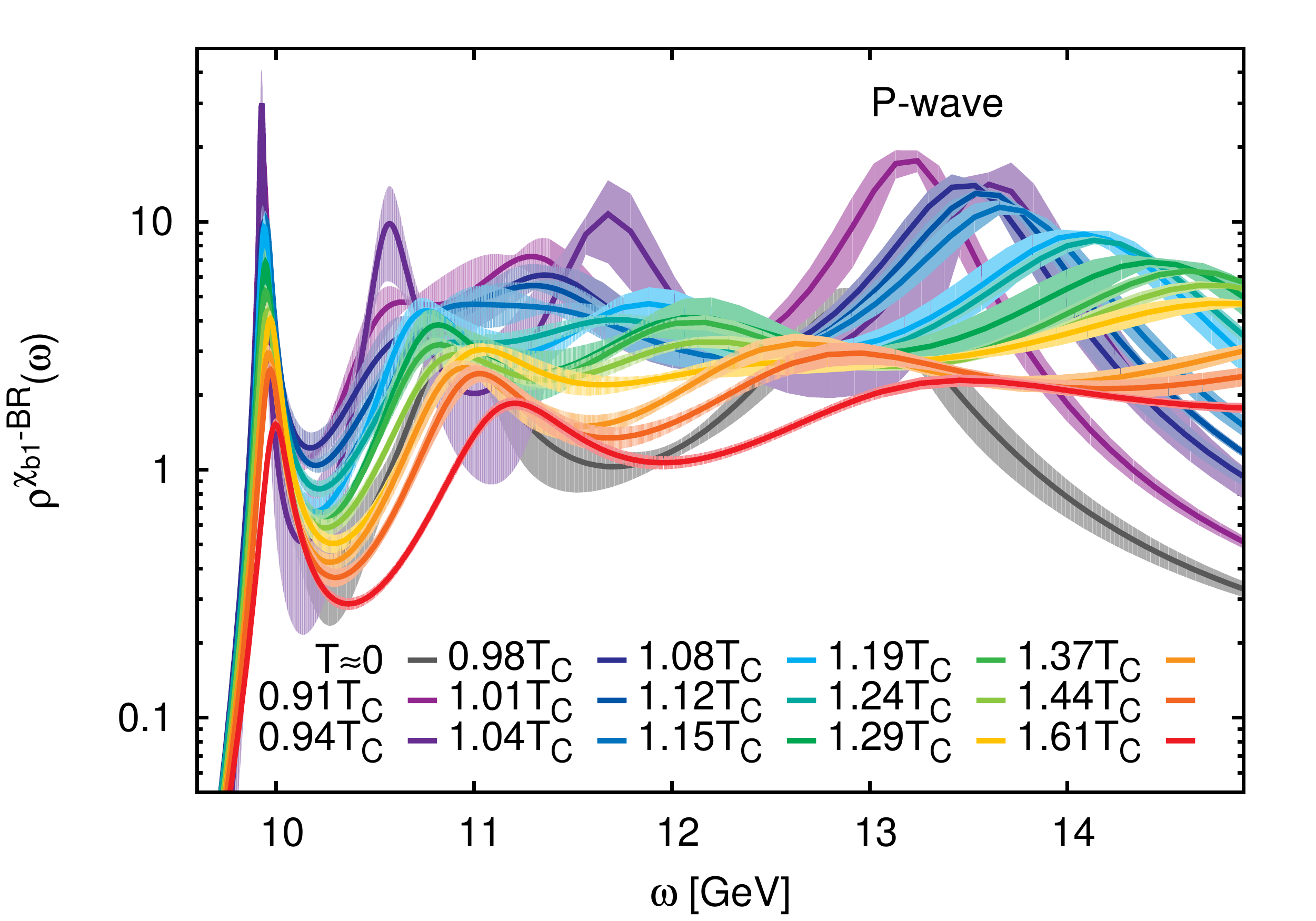}
 \includegraphics[scale=0.3, angle=-90]{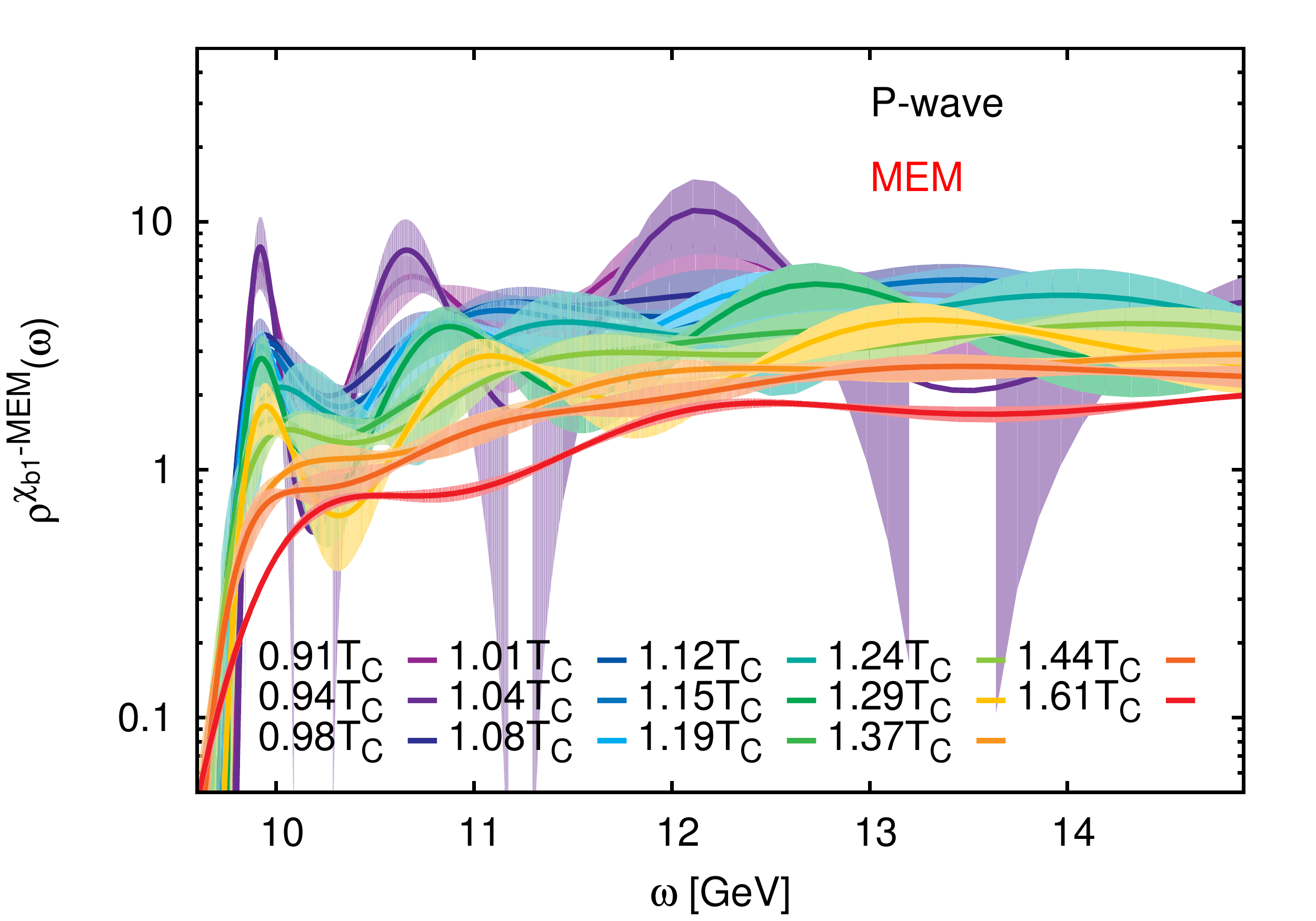} 
 \caption{P-wave bottomonium spectral functions at finite temperature from the novel Bayesian approach (left) and the MEM (right). The fourteen $T>0$ curves span the range of $140$MeV$<T<249$MeV. Again the Bayesian reconstruction shows a well resolved and Lorentzian shaped ground state peak at all temperatures. In the MEM results on the other hand the ground state feature is lost at $T\simeq211$MeV $(1.37T_c)$. }\label{Fig6}
\end{figure}

The spectra for the P-wave channel are shown in Fig.\ref{Fig6}. The left panel contains the results from the novel Bayesian approach, the right panel that from the MEM. As expected from the smaller signal-to noise ratio in the underlying correlators, the reconstruction is less reliable than in the S-wave case. Interestingly the two approaches in this case show even qualitatively different outcomes. While the novel Bayesian approach is able to resolve a well defined ground state peak even at the highest temperature of $249$MeV, the corresponding structure disappears in the MEM at around $211$MeV ($1.37T_c$). Note that while the changes in mass and width of the ground state at different temperatures, extracted from the Bayesian reconstruction, are easily distinguishable by eye, they are still smaller than the accuracy limits obtained for the P-wave channel at $T=0$.


To arrive at a definite statement about the survival or melting of the ground state, in particular in the P-wave channel, a simple inspection by eye is not sufficient. E.g. at high temperatures, the first and second peaked feature in the left panel of Fig.\ref{Fig6} take on the same amplitude. Without additional input we cannot decide whether a bound state persists or not. Here we propose to give a solid footing to the determination of melting through the comparison to free spectral functions, reconstructed from non-interacting NRQCD lattice correlators (for details see \cite{Kim:2014iga}). While obviously no bound states are encoded in the free lattice correlators, the spectral reconstruction might contain wiggly features due to the finite number of datapoints available. Comparing to the fully interacting spectra allows us to distinguish between this numerical Gibbs ringing and actual bound state peaks encoded in the finite temperature lattice data.

\begin{figure}[t]
\centering
 \includegraphics[scale=0.3, angle=-90]{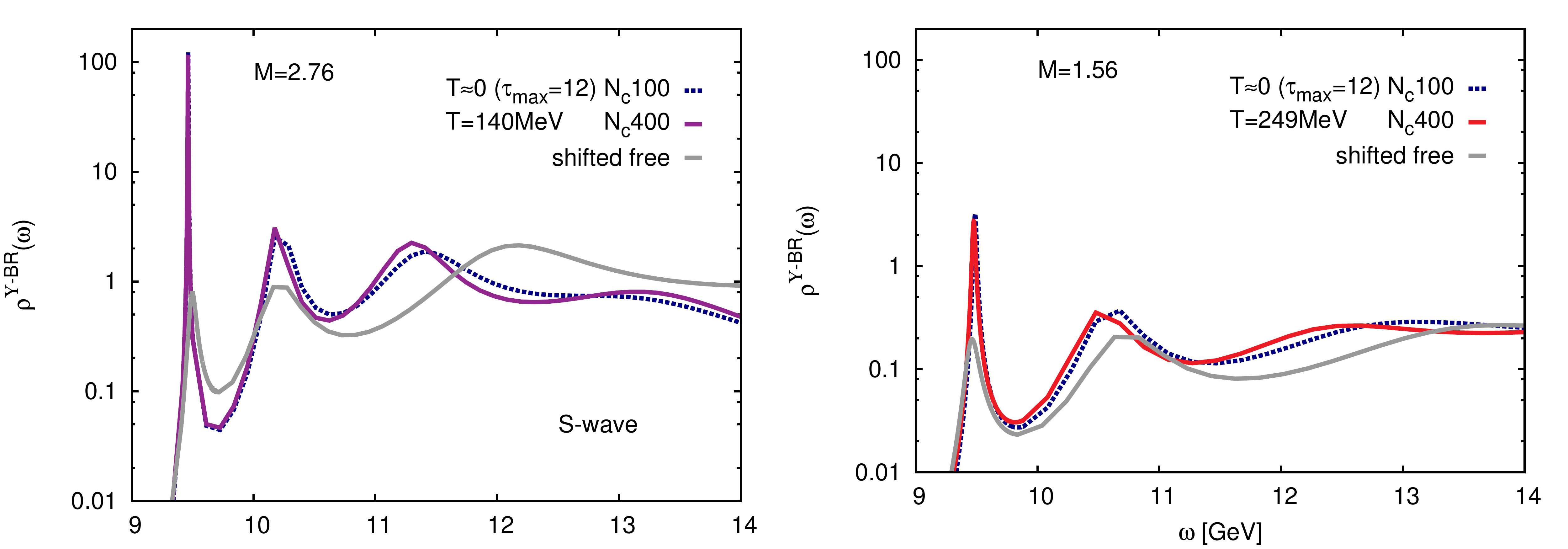}
 \caption{Comparison of the reconstructed S-wave spectra from interacting (colored solid lines) and non-interacting (solid gray lines) lattice NRQCD correlators at low ($140$MeV, left) and high ($249$MeV, right) temperature. The colored dashed curve denotes the result from the truncation of the $T\simeq0$ correlator dataset at the same lattice spacing. Note that numerical ringing is present in the free spectra, which however can be clearly distinguished from a bound ground state signal in the interacting spectra.}\label{Fig7}
\end{figure}

\begin{figure}[t]
\centering
 \includegraphics[scale=0.3, angle=-90]{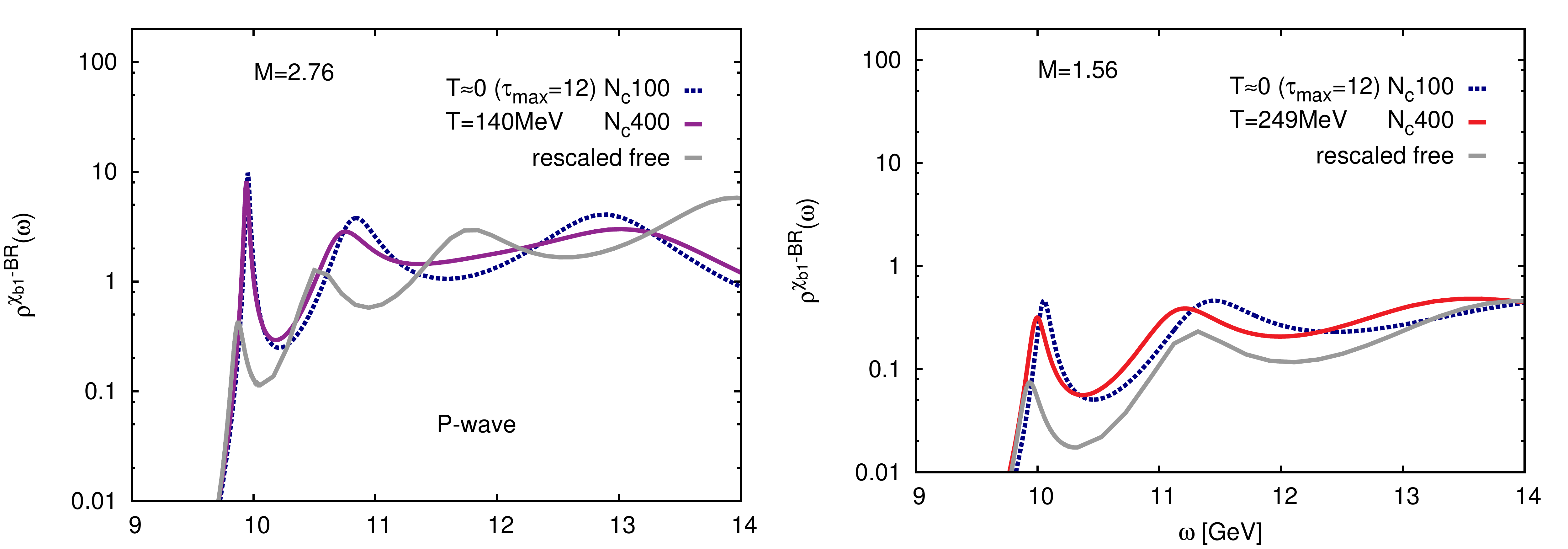} 
 \caption{Comparison of the reconstructed P-wave spectra from interacting (colored solid lines) and non-interacting (solid gray lines) lattice NRQCD correlators at low ($140$MeV, left) and high ($249$MeV, right) temperature. The colored dashed curve denotes the result from the truncation of the $T\simeq0$ correlator dataset at the same lattice spacing. While the difference between free and interacting spectra is smaller than in the S-wave case, even at $T=249$MeV the ground state signal remains a factor of three larger than the numerical ringing.}\label{Fig8}
\end{figure}

Fig.\ref{Fig7} and Fig.\ref{Fig8} show this comparison for the S-wave and the P-wave channel respectively. On the left we plot the lowest temperature ($140$MeV), on the right the highest temperature ($249$MeV). Each panel contains three curves. The colored solid curve denotes the finite temperature spectrum, while the colored dashed curve represents the reconstruction that one obtains from truncating the $T=0$ correlator dataset to $\tau_{\rm max}/a=12$ points. The gray solid curve depicts the non-interacting spectrum, which we have shifted by hand to cover the same frequency range as the interacting ones. For the S-wave, the distinction between numerical ringing and bound state signal is clear, both at $T=140$MeV and $T=249$MeV. At low temperatures the P-wave also shows an enhancement of more than one order of magnitude of the bound state signal over the numerical ringing. Even though the strength is smaller at $T=249$MeV, where it amounts merely to a factor of three, ringing and bound state state signal can be distinguished. Based on these results we conclude that both the S-wave ground state $\Upsilon(1S)$, as well as the P-wave ground state $\chi_{b1}(1P)$ survive in the medium up to $T=249$MeV, the highest temperature investigated.


\section{Conclusion}

Lattice regularized NRQCD is a mature framework, suitable for the investigation of the in-medium properties of heavy quarkonium bound states. Combined with realistic $48^3\times 12$ gauge configurations with $N_f=2+1$ light flavors, provided by the HotQCD collaboration, we determined the spectra of S-wave ($\Upsilon$) and P-wave ($\chi_{b1}$) states at zero and finite temperature. The use of a novel Bayesian prescription to spectral function reconstruction allows us to obtain reliable results even on small lattices with $N_\tau=12$ extend. We can set stringent upper limits for the strength of the in-medium modification of the bottomonium states (see Eq.\eqref{Eq:SwaveBounds} and Eq.\eqref{Eq:PwaveBounds}) and find from a systematic comparison to free spectral functions that both $\Upsilon$ and $\chi_{b1}$ survive up to $T=249$MeV, the highest temperature available in our study.

%%%%%%%%%%%%%%%%%%%%%%%%%%%%%%%%%%%%%%%%%%%%%%%%
%% BACKMATTER
%%%%%%%%%%%%%%%%%%%%%%%%%%%%%%%%%%%%%%%%%%%%%%%%

\begin{theacknowledgments}
SK is supported by the National Research Foundation of Korea grant funded by the
Korean government (MEST) No.\ 2010-002219 and in part by
NRF-2008-000458. PP is supported by U.S.Department of Energy under
Contract No.DE-AC02-98CH10886. AR was partly supported by the Swiss
National Science Foundation (SNF) under grant 200021-140234.
\end{theacknowledgments}

\bibliographystyle{aipproc}   % if natbib is available


\end{document}